\documentclass[prl,twocolumn,preprintnumbers,tightenlines,superscriptaddress,nofootinbib]{revtex4}

\pdfoutput=1
\usepackage[english]{babel}
\usepackage{amsmath,amssymb,amsfonts,bm,bbm,slashed,subdepth}
\usepackage{graphicx}
\usepackage[sort&compress]{natbib}
\usepackage{xcolor}
\usepackage[normalem]{ulem}
\usepackage{hyperref}
\usepackage{cleveref}
\definecolor{red}{rgb}{1.0, 0, 0}
\usepackage{enumerate}
\usepackage{epsfig, subfigure}
\usepackage{setspace}
\usepackage{booktabs, tabularx}
\usepackage{units}
\usepackage{placeins}
\usepackage{multirow}
\usepackage{mathtools}

\newcommand{\DC}[1]{\textcolor{blue}{[DC: #1]}}
  
\newcommand{\be}{\begin{equation}}  
\newcommand{\ee}{\end{equation}}  
\newcommand{\bea}{\begin{eqnarray}}  
\newcommand{\eea}{\end{eqnarray}}  
\newcommand{\pd}[2]{\frac{\partial #1}{\partial #2}}
\newcommand{\pdd}[2]{\frac{\partial^2 #1}{\partial #2^2}}

\begin{document}

\preprint{IPMU19-0143}

\title{Solitosynthesis and Gravitational Waves}

\author{Djuna Croon}
\email{dcroon@triumf.ca}
\affiliation{TRIUMF, 4004 Wesbrook Mall, Vancouver, BC V6T 2A3, Canada}

\author{Alexander Kusenko}
\affiliation{Department of Physics and Astronomy, University of California, Los Angeles\\
Los Angeles, California, 90095-1547, USA}
\affiliation{Kavli Institute for the Physics and Mathematics of the Universe (WPI), UTIAS\\
The University of Tokyo, Kashiwa, Chiba 277-8583, Japan}

\author{Anupam Mazumdar}
\affiliation{Van Swinderen Instituut, Rijksuniverseit Groningen, the Netherlands}

\author{Graham White}
\email{gwhite@triumf.ca}
\affiliation{TRIUMF, 4004 Wesbrook Mall, Vancouver, BC V6T 2A3, Canada}

\begin{abstract} 
We study the gravitational wave phenomenology in models of solitosynthesis. In such models, a first order phase transition is precipitated by a period in which non-topological solitons with a conserved global charge (Q-balls) accumulate charge. As such, the nucleation rate of critical bubbles differs significantly from thermal phase transitions. In general we find that the peak amplitude of the gravitational wave spectrum resulting from solitosynthesis is stronger than that of a thermal phase transition, while the timescale of the onset of nonlinear plasma dynamics may be comparable to Hubble. We demonstrate this explicitly in an asymmetric dark matter model, and discuss current and future constraints in this scenario. 
\end{abstract}

\maketitle

\section{Introduction}
Solitosynthesis~\cite{Griest:1989bq} of Q-balls may result in a first order phase transition of a distinct kind \cite{Kusenko:1997hj,Postma:2001ea,Pearce:2012jp}.  
Q-balls, carrying a global charge, may develop a scalar field condensate in their interiors, effectively lowering their free energy~\cite{Coleman:1985ki}. In a false vacuum, the scalar vacuum expectation value inside a Q-ball can reach the vicinity of the true vacuum. In this case, Q-balls can grow through the accretion of global charge due to solitosynthesis (a process similar to nucleosynthesis) until a critical size is reached. At this size the available free energy drives them to expand, completing the phase transition to the true vacuum.
This kind of a phase transition differs from a transition by tunneling, in which  a critical bubble of the true vacuum appears  due to quantum or thermal fluctuations~\cite{Kobzarev:1974cp,Coleman:1977py,Linde:1981zj}.  Instead, 
sub-critical bubbles stabilized by a conserved global charge form and grow gradually, until they reach the critical charge. Therefore, phase transition driven by solitosynthesis may be efficient in theories in which thermal tunneling is suppressed. 

This kind of a phase transition relies on a conserved global charge with a net asymmetry, and has therefore been studied in the context of minimal supersymmetric Standard Model (MSSM) \cite{Kusenko:1996jn,Pearce:2012jp}, which abounds with scalar fields carrying the baryon and lepton number; for a review see~\cite{Enqvist:2003gh,Dine:2003ax}.  The same process has a natural implementation in models of asymmetric dark matter \cite{Zurek:2013wia,Petraki:2013wwa,Oncala:2018bvl}.~\footnote{Stable Q-balls have also been considered as a dark matter candidates \cite{Kusenko:1997si,Kusenko:1997vp,Ponton:2019hux}.}

The gravitational wave phenomenology of first order phase transitions has recently received much attention (for a review see~\cite{Mazumdar:2018dfl}).
Bubble growth and coalescence source plasma dynamics, which in turn leads to dissipation of some of the released energy as gravitational radiation.
The resulting stochastic background spectrum depends solely on a few thermodynamic parameters: the temperature at which bubbles nucleate (or coalesce), the rate at which they nucleate, the velocity with which the bubble walls expand, and the amount of energy released to the surrounding plasma. For this reason, the phenomenology of different microphysical models may be very similar. 

In thermal phase transitions, the nucleation probability and critical size of a stable bubble are determined by the tunneling action. In contrast, critical bubbles in solitosynthesis are formed through the accretion of charge in thermal equilibrium. This difference implies that the effective nucleation rate for solitosynthesis may be much smaller. This has several implications for the gravitational wave phenomenology.~\footnote{The formation of Q-balls is also associated with gravitational waves from the fragmentation of the initial condensate~\cite{Kusenko:2008zm,Kusenko:2009cv}.}
The small nucleation rate implies that the phase transitions are typically very supercooled, and the released latent heat is large. Interestingly,  the effective nucleation rate may remain small for large latent heat, delaying the onset of nonlinear motion in the plasma. In particular, the suppression studied in \cite{Ellis:2018mja} and earlier mentioned in \cite{Hindmarsh:2017gnf} does not always apply. 

In this paper we explain the necessary conditions for the formation of Q-balls, and show explicit examples of potentials for which solitosynthesis is efficient. We then derive how to calculate thermal parameters needed for gravitational wave phenomenology, before performing calculations in an explicit example of an asymmetric dark matter model.

\section{Q-balls and Solitosynthesis}\label{sec:solito}
Scalar fields which carry a charge under a global symmetry can form coherent lumps named Q-balls in the presence of a primordial charge asymmetry \cite{Coleman:1985ki}. Once formed, Q-balls are stable due to charge conservation. 
The energy of a Q-ball with charge $Q$ and radius $R$ in the thin wall approximation is given by \cite{Kusenko:1997zq,Kusenko:1997hj,Postma:2001ea,Pearce:2012jp}
\bea \label{eq:thinwallE}
    E(Q,T)&=& \notag - | \Delta V(T)|\left( \frac{4 \pi}{3} R^3  \right) \\ &&+ 4 \pi R^2 S_1   + \frac{Q^2}{2 v(T)^2 (4/3) \pi R^3} \ .
\eea
where $ \Delta V(T)$ is the difference in the thermal scalar potential $V(\phi ,T)$ between the true and false vacuum, evaluated at temperature $T$, 
\begin{equation} S_1={\rm Re}\left[\int _0 ^{v(T)} \sqrt{2  V(\phi ,T)} d\phi\right]\end{equation} is the surface tension of the bubble and $v(T)$ is the field value of the true vacuum for a given temperature.
In chemical equilibrium, we study the growth of Q-balls as a function of the temperature. We assume that the microscopic processes involved are fast compared to the expansion of the Universe. In particular we require that the charge capture rates are larger, such that the production and growth of Q-balls does not undergo freeze-out.
With these assumptions, the number density of Q-balls of a given (conserved) charge is given by,
\begin{eqnarray} \notag
    n_Q &=& \frac{g_Q}{g_\phi ^Q} n_\phi ^Q \left( \frac{E(Q,T)}{ m_\phi } \right) ^{3/2} \left( \frac{2 \pi }{m _\phi T} \right)^{3(Q-1)/2} e^{ B_Q/T } \\ 
    n_\phi &=& \eta _\phi n _\gamma - \sum _Q Q n_Q, \label{eq:soliton} 
\end{eqnarray}
where $B_Q \equiv Q \,m_\phi -E(Q,T)$, $n_\gamma$ is the photon density, and $\eta _\phi$ is the charge asymmetry and $n_\phi $ is the number density of the particle carrying the global charge. Finally, $g_\phi$ indicates the number of degrees of freedom of $\phi$, and $g_Q$ is internal partition function of the Q-ball. In the following, we will use $g_Q =1$.
Tracking the population of Q-balls of charge $Q$ involves simultaneously solving Eqs.(\ref{eq:thinwallE}) and (\ref{eq:soliton}).

Bubbles which accumulate a critical charge expand due to kinematic pressure, and can reach very large wall velocities $v\sim 1$ before collision. This critical charge is defined as the threshold charge for which it becomes energetically favourable for the Q-ball to expand purely due to pressure,
\be
\left. \pd{E}{R} \right|_{Q=Q_c}=0, \,\,\,\,\,\,\,\,\,\,\,\,\,\,\,\,\,\,\,\,\, \left. \pdd{E}{R} \right|_{Q=Q_c}=0, \label{eq:critcon}
\ee
which, given \eqref{eq:thinwallE} implies,
\begin{equation}
    Q_C \approx \frac{100 \sqrt{10} }{81 } \frac{v(T)S_1^3}{| \Delta V(T)|^{5/2}} \ 
\end{equation}
in the thin-wall limit. Let us study this expression before solving  Eqs.(\ref{eq:critcon}) numerically in the following. First, we note that $S_1$ decreases while $ \Delta V(T)$ increases as the Universe cools. This means that the critical charge decreases with temperature, and stable sub-critical bubbles will eventually grow explosively. Furthermore, the population of large Q-balls is controlled by the factor $\exp (B_Q/T)$, implying that the population of large Q-balls also grows inversely temperature. By both effects, then, the effective nucleation rate of critical Q-balls will grow as the temperature decreases. The subsequent evolution and collision of Q-balls will not differ from that of bubbles in a transition mediated by thermal tunneling. This allows us to adopt the lattice results for the gravitational wave phenomenology, with an effective nucleation rate defined by Eq.\eqref{eq:soliton}. \par

By contrast, the thermal tunneling rate for bubble nucleation is controlled by the effective action normalized to the temperature, $\Gamma \sim T^4 \exp{(-S_E/T)}$. Here the Euclidean tunneling action $S_E$ is evaluated on the solution to the Euler Lagrange equations for the scalar field, $\phi$, which varies continuously from one vacuum to another \cite{Coleman:1977py}.  
The minimum of $S_E/T$ tends to be near the critical temperature, such that thermal tunneling typically takes place before the solitosynthesis nucleation rate becomes large. 
But in the case of a supercooled transition, the minimum of $S_E/T$ can be so large that bubble nucleation through tunneling is effectively suppressed throughout the thermal evolution.
This is what characterizes the processes we study below. 

We will give three categories of potentials that allow for a phase transition precipitated by solitosynthesis. The first kind has additional scalar fields at a similar mass scale. For example, in the MSSM augmented by a real scalar field, 
triscalar couplings are permitted between stops/sleptons and Higgs fields \cite{Kusenko:1996jn,Profumo:2007wc,Profumo:2014opa,Beniwal:2018hyi}. 
In this case, if one makes the appropriate rotation 
there exists a direction in field space with a cubic term. Then, we may parameterize the potential in this direction as follows \cite{Croon:2018erz}
\begin{equation}
V(\phi ) = \Lambda ^4 \left( \left[ \frac{3-4  a }{2}  \right] \left( \frac{\phi}{v_\phi} \right)^2- \left( \frac{\phi}{v_\phi} \right)^3 +  a  \left( \frac{\phi}{v_\phi} \right)^4 \right) \label{eq:pot1}   \ .
\end{equation}
The physical scales are inputs to the potential, and we have parametrized the model such that a tree level barrier, which may lead to supercooling, exists for $1/2<{ a} <3/4$. \par 
A second category of supercooled potentials extensively studied in the literature arises when a heavy field is integrated out, producing a dimension-6 interaction in the scalar potential. Fermionic and bosonic loop contributions can produce negative correction to the quartic interaction in the effective theory at low energies \cite{Grojean:2004xa,Corbett:2017ieo,Delaunay:2007wb,Chala:2018ari,Ellis:2018mja,Ellis:2019flb,Ellis:2019oqb} and a relative sign difference between the sextet and quartic. Thus substantial supercooling can occur. 
We can again parametrize the low energy effective potential with the physical scales as inputs \cite{Croon:2018erz}
\begin{equation}
    V(\phi ) = \Lambda ^4 \left( \left[ 2-3 a \right] \left( \frac{\phi}{v_\phi } \right) ^2- \left( \frac{\phi}{v_\phi } \right) ^4 +a  \left( \frac{\phi}{v_\phi } \right) ^6 \right)  \ . \label{eq:moneypot}
\end{equation}
In this case a tree level barrier exists for $1/2 <  a < 2/3$. \par 
A final type of supercool potential, arises from conformal symmetry breaking \cite{Jinno:2016knw,Marzo:2018nov,Ellis:2019oqb}. In this case the dominant term in the potential is from the $\beta $ function (typically from a gauge field) which in the presence of an effective thermal mass implies a thermal barrier, 
\begin{equation}
    V(\phi ,T) \sim \beta A \phi^4 \left( \log \left[ \frac{\phi^2}{v_\phi ^2}\right] - C \right) + c_T T^2 \phi ^2. \label{eq:pot3}
\end{equation}
Here the coefficients $A$, $C$ and $c_T$ are model-dependent numerical factors. For the remainder of this paper we will focus on potentials of the form in Eq.\eqref{eq:moneypot}, in a benchmark asymmetric dark matter model. However, we note that our analysis can easily be generalized to any potential that leads to supercooling. We leave the analysis of these other potentials to future work.

\section{Gravitational waves from Solitosynthesis}\label{sec:GW}
If a first order phase transition occurs in a plasma, the latent heat that it releases may be transferred to the degrees of freedom coupled to the bubble wall. This is known to give rise to acoustic waves in the plasma, which sources gravitational waves for a period after the completion of the transition \cite{Hindmarsh:2016lnk}. In this section we will describe the calculation of the thermal parameters which govern the dynamics of the phase transition and the resulting gravitational wave spectrum. 

We will focus primarily on the nucleation rate of Q-balls with the critical charge, noting that the bubbles with sub-critical charge typically occupy a very subdominant volume fraction at the collision temperature and can therefore be ignored in the gravitational wave calculation. In analogy with thermal phase transitions, we may parametrize the nucleation rate as $\Gamma(\tau) = \Gamma_f \,\text{Exp}(\beta (\tau-\tau_f))$ in terms of conformal time \cite{Caprini:2007xq}. This defines the parameter $\beta$, 
\begin{equation}
    \frac{\beta }{H_*}= T_* \frac{\dot{ \Gamma } }{\Gamma } \ ,
\end{equation}
which is normalized to the Hubble rate,  $H_*$, as is conventional, and the subscript $*$ implies that the quantity should be evaluated at percolation. In solitosynthesis, we may calculate $\Gamma _Q$ for a Q-ball with charge $Q$ from \eqref{eq:soliton} and the relation \cite{Caprini:2011uz}, 
\begin{equation}
\frac{4 \pi }{3 } n_{Q} (T) H^{-3}(T)  = \int _{T_C}^T \frac{1}{H(\bar{T})\bar{T}} V_H(\bar{T}) \Gamma _Q (\bar{T} ) d \bar{T}  \ ,
\end{equation}
where $V_H(T)\equiv 4 \pi H^{-3}(T)/3$ is the Hubble volume.
Q-balls with the critical charge expand near-relativistically, 
as bubbles in thermal phase transitions do. 
Finally, the volume fraction occupied by the Q-balls is $1-e^{-f(T)}$, with
\begin{equation}
    f(T) = \sum _Q \frac{4 \pi }{3} \int _T ^{T_C} \frac{d \bar{T}}{\bar{T} H^4(\tilde{T})} \Gamma _Q (\bar{T} ) \left( \int _T ^{\tilde{T}} \frac{d \tilde{T}}{H(\tilde{T})} \right)^3 \ .
\end{equation}
We define the percolation temperature $T_\ast$ at $f(T_\ast)=1$. Because of the near-relativistic growth of critical bubbles, their radius exceeds that of sub-critical bubbles by many orders of magnitude at collision.  Therefore in defining the inverse time scale we can make the approximation $\Gamma \sim \Gamma _{Q_C}$.
Importantly, the latent heat $\alpha$ can be found in analogy with a thermal phase transition, 
\begin{equation}
    \alpha = \frac{\Delta V - \frac{1}{4}T d\Delta V/dT}{\rho _{\rm rad}} \Bigr\rvert_{T = T_\ast}
\end{equation}
where $\rho _{\rm rad}$ is the radiation energy density.
 
Although the microphysics describing the thermal parameters differs from that of a thermal phase transition, the relationship between the thermal parameters and the gravitational wave spectrum today is the same. For the models considered here, this spectrum is dominated by the acoustic plasma motion \cite{Hindmarsh:2017gnf,Weir:2017wfa}
\begin{eqnarray}
h^2 \Omega _{\rm sw} (f) &=& 8.5 \times 10^{-6}  \\ &\times& \left( \frac{100}{g_\ast}  \right)^{1/3} \kappa _f^2 \frac{\alpha ^2}{(1+\alpha)^2} \left( \frac{H_\ast}{\beta} \right) v_w S_{\rm sw} (f)H_\ast t_{\rm sw}\nonumber 
\end{eqnarray}
with a peak frequency given by
\begin{equation}
f_{\rm sw} = 6.2 \, \mu {\rm Hz} \, \left(\frac{1}{v_w}\right) \left( \frac{\beta}{H_\ast} \right) \left( \frac{T_\ast}{100{\rm GeV}} \right) \left( \frac{g_\ast}{100} \right) ^{1/6} 
\end{equation}
and the timescale on which acoustic waves are active can be estimated from \cite{Ellis:2019oqb}
\begin{equation}
H_\ast t_{\rm sw} = {\rm Min} \left[1 , \frac{2(8 \pi )^{1/3} v_w \sqrt{1+\alpha }}{\sqrt{3\alpha \, \kappa _{f}}\, \beta/H_\ast } \right] \ .
\end{equation}
The fraction of vacuum energy that is converted to kinetic energy in the plasma for ultra-relativistic walls is given by \cite{Espinosa:2010hh} 
\begin{equation}
\kappa _f = \frac{\alpha }{0.73 + 0.083\sqrt{\alpha } +\alpha } \ . 
\end{equation}
Finally, the spectral form $S_{\rm sw}(f)$ is a broken power law, the form of which can be found in ref. \cite{Hindmarsh:2017gnf,Weir:2017wfa}.

Let us conclude this section by noting that the Hubble constant includes contributions from the potential energy in the false vacuum. When the vacuum energy dominates over the radiation energy in the plasma, the above analysis becomes invalid, as the Q-ball number density no longer obeys the Saha equation \eqref{eq:soliton}. We leave the analysis of phase transitions preceded by solitosynthesis during vacuum domination to future work. 

\section{Solitosynthesis in an asymmetric dark sector}
Asymmetric DM (ADM) models feature a hidden sector with a conserved global symmetry, under which the Universe has a net charge. ADM models also include an interaction which annihilates away the symmetric dark plasma, which motivates a hidden gauge structure. Hence, we may study solitosynthesis in a hidden sector.\footnote{For a recent review on gravitational wave phenomenology of dark sectors, see \cite{Bertone:2019irm}.}

Assuming a thermal production mechanism, the DM yield is proportional to the charge asymmetry. Then, the DM mass and asymmetry are related by \cite{Petraki:2013wwa},
\bea
\frac{m_{\rm DM}}{m_p}\, \frac{\eta_{\rm DM}/q_{\rm DM}}{\eta_B} \,\frac{1+r_{\infty}}{1-r_{\infty}}= \frac{\Omega_{\rm DM}}{\Omega_{\rm SM}} \approx 5,
\eea
where $\eta_{\text{DM},B}$ are the dark and baryon asymmetry respectively, $m_{\rm DM}$ and $q_{\rm DM}$ are the mass and charge of the DM particle, and $r_\infty$ is the fractional asymmetry of the dark sector at late times. In the following, we will assume minimal models with $q_{\rm DM} =1$ and a fully asymmetric dark sector, such that $r_\infty = 0$. Then, a large asymmetry implies that the lightest dark particle has a mass,
\be \label{eq:massabun}
m_{\rm DM} \approx \left(\frac{\eta_{\rm DM}}{10^{-3}}\right)^{-1} 3 \,\, \text{keV} .
\ee
We will assume that the dark matter today is asymmetric. The non-asymmetric components in the hidden sector are not protected by a global symmetry, and may therefore decay to particles in the visible sector. An example of a coupling which realizes this is a kinetic mixing between the hidden sector gauge boson and the SM photon $-\frac{1}{2}\kappa F_{\mu \nu} V^{\mu \nu}$. 

In a minimal model, the Q-balls are formed from the lightest scalar with a charge under the global symmetry. The scalar that forms the order parameter does not have a global charge, but is charged under the hidden sector gauge group. It may therefore decay, for example, to lighter hidden sector particles with fractional global charge (two fermions with ADM charge $1/2$, or three scalars with charge $1/3$), or to several lighter particles, one of which is charged.

\begin{figure}
    \centering
    \includegraphics[width=0.5\textwidth]{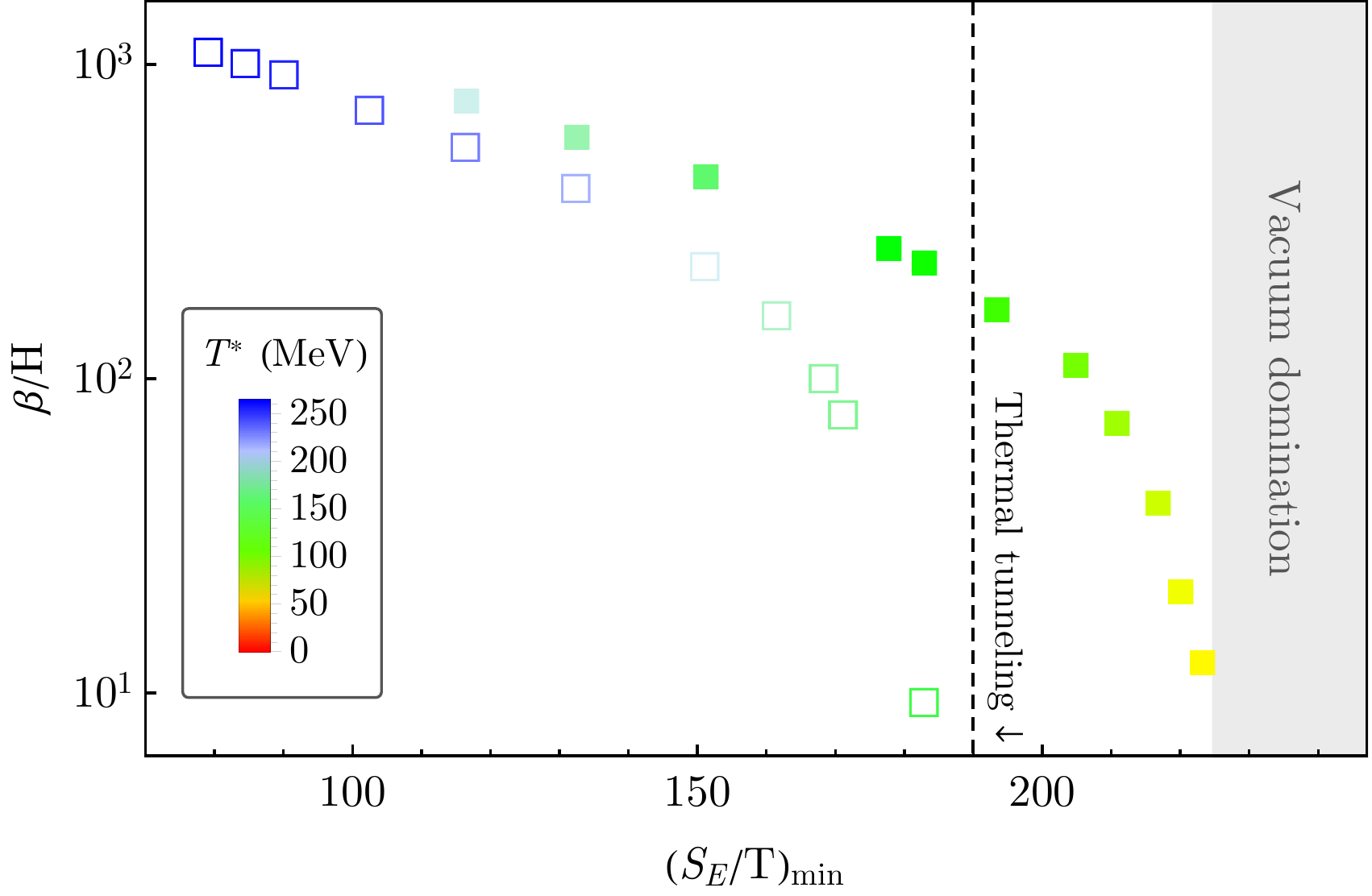}
    \caption{Benchmark study for $\eta  _{\rm DM}= 10^{-8}$, $v=50 $  MeV, $\Lambda =20$ MeV, and $g=0.1$. . Here, the filled squared indicate a phase transition triggered by solitosynthesis, and the open squared are for a thermal phase transition in the same model.}.
    \label{fig:SEbHexample}
\end{figure}

Let us denote the scalar charged under the new gauge group $\phi$ and the scalar with a global charge $\varphi$. The model is then described by the Lagrangian
\begin{equation}
    {\cal L} \supset |D_\mu \phi |^2 + |\partial _\mu \varphi |^2   - V(\phi,\varphi)  -\frac{1}{2}\kappa F_{\mu \nu} V^{\mu \nu}
\end{equation}
where the mixed quartic coupling is given by $y |\phi|^2 |\varphi |^2 $ and the $\varphi$-self-coupling by $\lambda  |\varphi^\dagger \varphi| ^2/4$ (note that the mixed quartic coupling will induce an effective cubic term at finite temperature). For the potential of $\phi$ we can use any potential the forms given in Eqs. (\ref{eq:pot1}--\ref{eq:pot3}). We choose one of the form \eqref{eq:moneypot} and include finite temperature corrections.
 These corrections depends upon the $\phi$-dependent masses of the gauge boson, the dark matter candidate that carries the asymmetry as well as the physical mass of $\phi$ and the Goldstone mode (for a review see \cite{Mazumdar:2018dfl}).

To admit Q-balls, the total scalar potential $ V(\phi,\varphi,T)$ must satisfy  the usual condition~\cite{Coleman:1977py}, namely,
\begin{equation}
    V(\phi_0,\varphi,T)/\varphi^2 =\min \ {\rm for}\ \varphi=\varphi_0\neq 0. 
\end{equation}

Note that this condition is automatically satisfied for any positive value of $\lambda$.\footnote{ This can most easily be understood in the high temperature expansion, which contains an effective negative term $\propto -( \partial^2V/\partial\phi^2 + y |\varphi|^2)^{3/2} T$. One can verify numerically that the thermal corrections away from this limit also allow for the existence of Q-balls.}
We assume that the Coleman Weinberg contribution redefines the zero-temperature parameters in Eq.\eqref{eq:moneypot} and do not consider it explicitly. The mixed quartic coupling $y$ is set by the relic abundance of the scalar dark matter candidate (see Eq.\ref{eq:massabun}). The only other free parameter is then the gauge boson coupling, $g$. We assume the hidden sector is self thermalized but remain agnostic about whether it is in thermal contact with the standard model degrees of freedom during the phase transition. The only change in such a case would be to mildly shift the temperature at which vacuum domination occurs as well as a mild suppression in the latent heat. For concreteness we assume no thermal contact, and that the temperature of the hidden sector is approximately equal to the temperature of the visible sector.

Let us now motivate our benchmark choices. Thermal tunneling is suppressed and the gravitational wave signal is strengthened for a sizeable ratio of $x=v/\Lambda $ \cite{Croon:2018erz}. However, the hidden sector gauge boson must be lighter than $m_\phi$ to deplete the symmetric part of the hidden sector, such that we choose a value of $x=2.5$. A simple numerical check verifies that the effective Wilson coefficient of the dim-6 operator is sufficiently small to motivate our EFT treatment. 
We fix the mixed quartic coupling between the asymmetric scalar and $\phi$ by fixing the relic abundance to $\Omega_{\rm DM}$.
We study the regime of small gauge couplings, $g=10^{-1}$, in which the finite temperature formalism is valid,  mitigating poor infrared convergence \cite{Linde:1980ts}. Finally, we infer a minimum gauge boson mass from BBN constraints, which will in turn imply an approximate upper bound on the asymmetry~\footnote{With the caveat that a different decay mechanism, a hidden sector temperature that greatly differs from the visible sector, and a large gauge coupling could modify this argument.}. 

Assuming kinetic mixing to drain the hidden sector of its symmetric component, constraints on the mixing parameter $\kappa$ inform our choice of asymmetry $\eta _{ \rm DM}$. Supernova constraints enforce $\kappa \lesssim 10^{-10}$.
The neutron to proton ratio freezes out at $T=0.8$ MeV, below which hidden sector particles decaying into electromagnetic final states may in principle destroy light elements and imply entropy injections. The former is only relevant for very small kinetic mixing parameters, corresponding to lifetimes of $\tau > 10^4$s \cite{Poulin:2015opa,Forestell:2018txr,Hufnagel:2018bjp}. Following a procedure outlined in \cite{Forestell:2018txr}, we derive an upper bound on the asymmetry parameter $\eta \leq 10^{-7}$ from the upper bound on entropy injections for $m_V\lesssim 5$ MeV, assuming an initial thermal state. 
Informed by this upper bound, we study the benchmarks $\eta  _{\rm  DM} =10^{-7}$ and $\eta _{\rm  DM} =10^{-8}$. In the latter case, decays happen well before the onset of nucleosynthesis.

\begin{figure}
    \centering
    \includegraphics[width=0.45\textwidth]{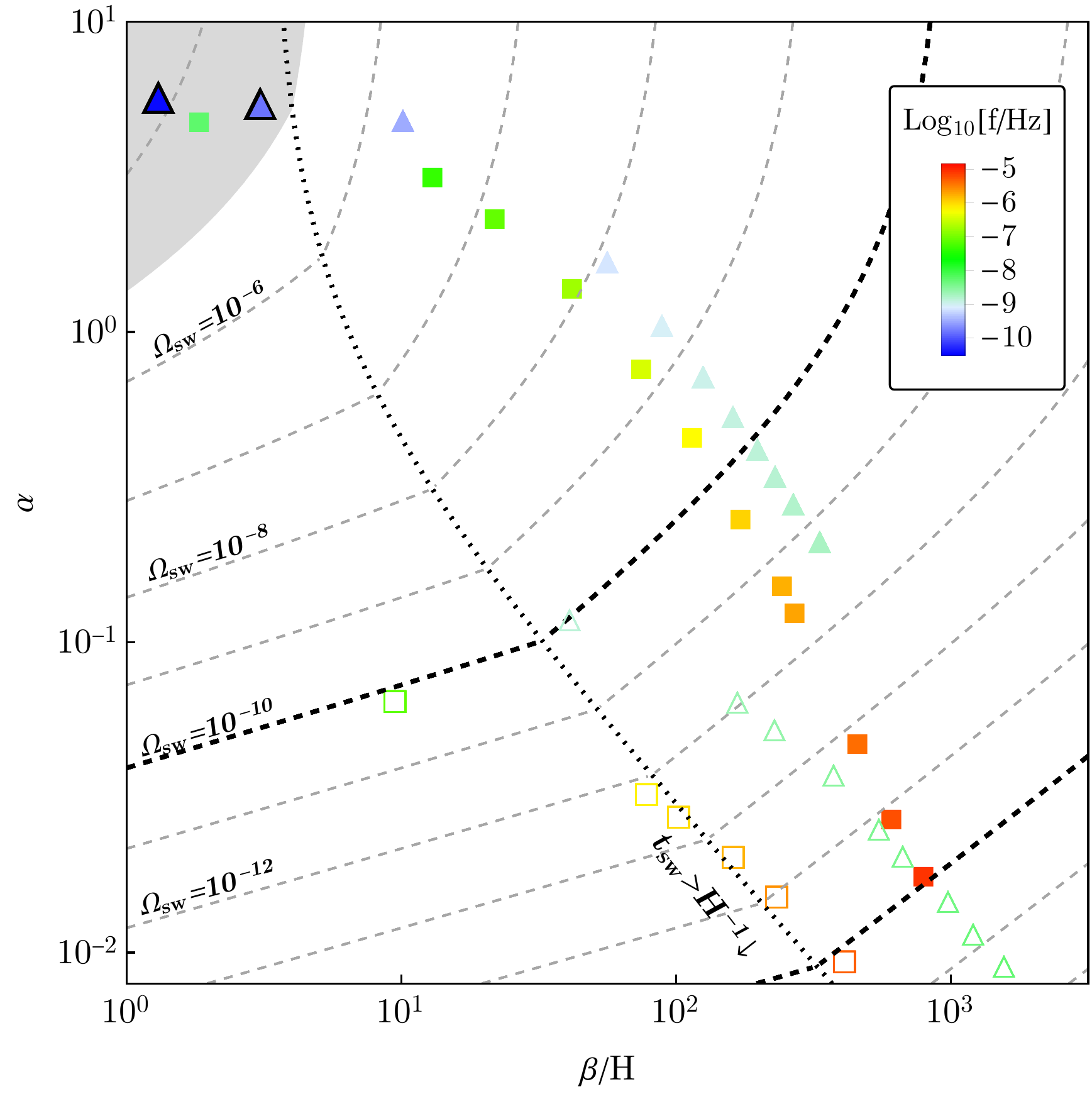}
    \caption{Predictions for the latent heat parameter $\alpha$ and transition rate $\beta/H$ in two benchmark cases, with $\eta  _{\rm DM}= 10^{-8}$ (squares) and $\eta _{\rm DM} = 10^{-7}$ (triangles). The filled figures are for phase transitions completed by solitosynthesis, and the open figures are for phase transitions that complete thermally. The benchmarks with a triangle frame are currently probed by Nanograv \cite{Arzoumanian:2018saf} and the grey shaded region indicates current constraints on $\Delta N_{\rm eff}$. The dashed lines give an estimate for the fractional GW density $\Omega_{\rm GW}$ which is taken from \cite{Hindmarsh:2017gnf}. The thicker dashed lines give the approximate peak sensitivities of the EPTA \cite{vanHaasteren:2011ni} and  SKA (projections) \cite{Moore:2014lga}. To the left of the dotted line, the formation of nonlinear dynamics in the plasma is assumed to be sufficiently slow for the acoustic waves to survive for a Hubble time.}
    \label{fig:GWs}
\end{figure}

We use \verb|Bubble-profiler| \cite{Athron:2019nbd,Akula:2016gpl} to find the thermal parameters in thermal phase transitions and the methods described in sections Eqs.(\ref{sec:solito} and \ref{sec:GW}) to find the nucleation and growth of the Q-balls, in the same model. To compare both nucleation processes, we vary the parameter $a$ in the potential Eq.\eqref{eq:moneypot}, which parametrizes the height of the barrier at zero temperature (and therefore the minimal ratio $S_E/T$).
We show the result of this calculation in Figs. \ref{fig:SEbHexample} and \ref{fig:GWs}. In the first of these figures, it is seen that solitosynthesis may indeed occur for phase transitions that would not complete through tunneling, as the smallest ratio $S_E/T$ is too large, indicating a nucleation rate smaller than Hubble. Secondly there is a correlation between $\beta/H_{ \ast}$ and the minimum value of $S_E/T$. 

In Fig.~\ref{fig:GWs}, we show the resulting thermal parameters and peak gravitational wave amplitude for the same benchmark, and a benchmark with a larger charge asymmetry. In this figures we allow a more liberal bound on vacuum domination, requiring that $H<2H_{\rm rad}$. We also show constraints from the frequency independent constraint on the amount of radiation ($\Delta N_{\rm eff}$) and which benchmark points are currently probed by the Nanograv experiment. While many of the other points have a large enough peak amplitude to be observed by pulsar timing arrays such as Nanograv, the fact that the peak frequency is proportional to $\beta / H$ renders many of our benchmark points outside their sensitivity window. Fig.~\ref{fig:GWs} demonstrates that the gravitational wave spectra from a phase transition precipitated by solitosynthesis are typically enhanced, in particular due to a large amount of supercooling associated with an increased latent heat. The typical time scales in such transitions are also longer. Note that the points with the largest latent heat also feature acoustic processes which last longer than the Hubble time.

\section{Conclusion}
In this paper we have studied the gravitational wave phenomenology of phase transitions triggered by solitosynthesis. We have described a procedure to calculate the thermal parameters for bubble nucleation through charge diffusion in a plasma. In a benchmark model of assymetric dark matter, we have compared the dynamics of solitosynthesis and tunneling, and found that the resulting gravitational wave spectrum is typically enhanced. This opens up the interesting possibility of probing asymmetric dark matter at (indirect) low frequency gravitational wave experiments such as pulsar timing arrays. 

{\bf Acknowledgements} TRIUMF receives federal funding via a contribution agreement with the National Research Council of Canada and the Natural Science and Engineering Research Council of Canada.  The work of A.K. was supported by the U.S. Department of Energy Grant No. DE-SC0009937 and by the World Premier International Research Center Initiative (WPI), MEXT Japan.  AM is supported by Netherlands Organisation for Scientific Research (NWO) grant no. 680-91-119.

\appendix 

\bibliography{references}
\end{document}